\documentclass[aps,prl,reprint,groupedaddress]{revtex4-2}

\usepackage{dcolumn}% Align table columns on decimal point
\usepackage{braket}
\usepackage{amsmath} %数学公式包，没有这个包敲公式可能出错
\usepackage{amsthm} % 数学定理包，提供写证明过程的环境的
\usepackage{graphicx} % Required to insert images
\usepackage{amsfonts}%解决dAlembert operator打不出的问题
\usepackage{float}     % 也是和图片设置有关的
\usepackage{physics} % 物理包，不知道好不好用，想打ket才找到这个的
\usepackage{subcaption}
\usepackage{booktabs}  % 绘制三线表用

\usepackage{xcolor}

\theoremstyle{definition}

\begin{document}

\title{ Symmetry-imposed correlation in nuclear level statistics: The spin distribution }

\author{ Junchao Guo }
\email[]{ guojunchao0704@sjtu.edu.cn }
\affiliation{School of Physics and Astronomy, Shanghai Jiao Tong University, Shanghai 200240, China}

\author{ Yang Sun }
\email[]{ sunyang@sjtu.edu.cn }\affiliation{School of Physics and Astronomy, Shanghai Jiao Tong University, Shanghai 200240, China}

\date{\today}

\begin{abstract}

Despite long-term research, the origin of spin cutoff in the angular-momentum (spin) distribution of nuclear level densities remains incompletely elucidated. We demonstrate that this problem can be traced back to Bethe’s assumption that nucleons in finite Fermi systems are independent random variables. By constructing a statistical ensemble that enforces rotational invariance through angular-momentum coupling, we obtain an analytical expression for the spin cutoff parameter, which includes a previously unidentified finite-population correction. Our results show that, even in the absence of interactions, nuclear many-body states exhibit non-negligible correlations arising from fermionic antisymmetry and angular-momentum coupling. From this perspective, spin cutoff may be interpreted as a quantitative measure of correlation imposed by symmetry in nuclear level statistics.

\end{abstract}

\maketitle

\textit{Introduction} -- When atomic nuclei are excited to sufficiently high energy, the average spacing between energy levels become too small to be resolved individually. At this point, the energy spectrum is considered continuous with statistical fluctuations \cite{Weidenmueller2009}.
This perspective is applicable within the framework of the Hauser-Feshbach (HF) reaction theory \cite{HF1952}, where the concept of the compound nucleus, first suggested by Niels Bohr
\cite{Bohr1936}, posits that after absorbing energy, the nucleus distributes the energy among many degrees of freedom. Subsequent $\gamma$-ray decay is statistical, governed by nuclear level density (NLD) and transmission coefficients, which forms the basis of modern reaction calculations.

The compound nucleus is often treated phenomenologically, where only conserved quantum numbers (energy $E$, angular momentum $J$, and parity $\pi$) are preserved. Therefore, the NLD required for reaction calculations should be a function of these quantum numbers. In his groundbreaking paper in 1936 \cite{Bethe1936}, Hans Bethe laid the foundation for calculating NLD based on the Fermi gas model \cite{Griffiths1995}. Bethe's derivation yielded the total number of nuclear levels  within given energy ranges, $\rho(E)$, independent of $J$ and $\pi$. To obtain the NLD's dependence on $J$ and $\pi$, the problem is decomposed into $\rho(E,J,\pi)= \rho(E) P(J) \Pi(\pi)$, where the functions $P(J)$ and $\Pi(\pi)$ are determined by other structural information outside the model.

Regarding angular momentum, determining $P(J)$ has been a key research focus. In the 1950s, Jensen and Luttinger \cite{Luttinger1952}, and Bloch \cite{Bloch1954}, attempted to derive the function from the Fermi gas model. In 1960, Ericson,  in an influential work \cite{Ericson1960}, derived a formula for $P(J)$. Following the same line of thought and employing various different methods, attempts have been made to determine the parameters in the formula \cite{Cameron1965,Cameron1969,Dilg1973,Su1991,Iljinov1992,Ignatyuk1993,Rauscher1997,Egidy2009,Grimes2016,Voinov2023}. Extensive effort has been devoted to empirically fitting the parameters in the energy-dependent $P(J)$ using experimental level schemes \cite{Egidy2009}. It can be concluded that although qualitative characteristics of spin distribution can be understood through simple  phenomenological or experimental methods, quantitative understanding remains a challenge \cite{Grimes2019}. 

Theoretical models capable of studying this problem must treat angular momentum as a good quantum number.  Nuclear shell models (exact or approximate) have been applied \cite{Johnson2001,Alhassid2005,Kaneko2007,Zelevinsky2016,Dang2017,Ormand2020}. A common problem is that shell-model calculations are usually limited in a small model space for light nuclei. In principle, the angular momentum projection method can handle large model spaces for heavy nuclei \cite{Sun2016}; numerical examples of level densities with detailed spin distribution have been demonstrated in the Shell Model Monte Carlo method \cite{Alhassid2007} and the Projected Shell Model \cite{Wang2023,Wang2025}. However, such calculations are often complex, making it difficult to clearly see the root of the problem.

The spin problem in level statistics has long plagued nuclear physicists. The lack of spin information in NLD severely impacts many applications, such as nuclear astrophysics \cite{Liddick2016}, fragment problems in fission theory \cite{Stetcu2014,Stetcu2013,Bulgac2021,Schunck2021}, and nuclear data evaluation \cite{Zeiser2019,Liu2024}. To identify the root, it is useful to recall Bethe's initial thought.

\vspace{0.2cm}
\textit{Recap of Bethe's initial idea} --  In his 1936 paper \cite{Bethe1936}, Bethe did not  explicitly write $P(J)$, but discussed how to derive it. Ericson's later work  \cite{Ericson1960} provided an explicit expression. Hereinafter, we will refer to this as Bethe-Ericson (BE) formula. 

Bethe started with a fundamental assumption that each random variable does not affect the others and that they come from the same probability distribution. This i.i.d. (independent and identically distributed) assumption is standard in probability theory and statistical modeling, leading naturally to results that conform to the Central Limit Theorem (CLT). For a nucleus with $N$ nucleons, the total $z$-component of angular momentum is 
\begin{equation}
\label{TotalM}
M= m_1+ m_2+ \cdots m_N.
\end{equation}
According to the CLT, for sufficiently large $N$, regardless of specific distribution of each $m_i$, as long as the variance is finite (which is obvious in the present context), the distribution of $M$ follows a Gaussian form. Furthermore, due to the symmetry with respect to the mean (which is zero), the distribution of $M$ takes the form
\begin{equation}\label{PM}
P(M) = \dfrac{1}{\sqrt{2\pi}\sigma} \exp(-M^2/2\sigma^2) ,
\end{equation}
where $\sigma$ is standard deviation of the $M$ distribution, and 
\begin{equation}\label{sigma2}
\sigma^2={\rm Var}(M)
\end{equation}
is variance of $M$.

To relate $P(M)$ to the distribution of total angular momentum $P(J)$, consider that for each $M$, $P(M)$ contains all $J\ge |M|$ contribution, i.e. 
\[
P(M) = P(-M) = \sum_{J\ge |M|} P(J) .
\]
In statistics, $1-P(M)$ is the so-called cumulative distribution function (CDF) of $P(J)$. %Take the difference we obtain
%\[
%P(M-1) - P(M) = P(J).
%\]
Assuming that $M$ (and $J$) can take very large values, the summation in the above equation can be replaced by an integral, and then $P(J)$ can be obtained by differentiating the CDF, i.e.
\[
P(J) \propto -\dfrac{dP(M)}{dM}|_J .
\]
The differentiation leads to the well-known BE formula
\begin{equation}\label{BE}
P(J) = \dfrac{2J+1}{2\sigma^2}\exp[-\dfrac{J(J+1)}{2\sigma^2}] .
\end{equation}

\vspace{0.2cm}
\textit{Discussion of the BE formula} --
Equation (\ref{BE}) leaves $\sigma$ undefined. This single parameter absorbs {\it all} nuclear structure effects and determines how rapidly the exponential suppresses high-spin states. Therefore, $\sigma$ is commonly referred to as {\it spin cutoff} parameter.

The derivation of Eq. (\ref{BE}) was based on the assumption that each nucleon is independently and identically distributed, and that the number of nucleons contributing to the total angular momentum is large. Under these conditions, the generation of many-body angular momentum is simplified to a single-particle problem, where the total angular momentum arises from the random coupling of the angular momenta of individual particles \cite{Ericson1960}.

However, nuclei are finite quantum systems. If excitation energy  (temperature) is not too high, the assumption in \cite{Ericson1960} tends to overestimate statistical fluctuations. In fact, in many practical cases, nucleon occupation is mainly confined in a few $j$-shells, and the way for them to form total angular momentum cannot be considered a random process. In our recent work \cite{Guo2025}, we proved that nucleons within a single $j$-shell exhibit maximal correlations caused by angular momentum coupling. These correlations lie in Hilbert space and always exist in nuclei that respect rotational symmetry. This means that nucleons cannot be considered independent.

In this Letter, we show that enforcing rotational invariance while taking Pauli exclusion principle in a statistical ensemble fundamentally alters the structure of nuclear many-body states. We can analytically derive the function $P(J)$ and give an expression for the spin-cutoff parameter $\sigma$ without relying on external quantities.

\vspace{0.2cm}
\textit{ Our method} --  We first recall some elements of Fermi-Dirac statistics relevant to this work. For a set of single particles with energy $\varepsilon_i$, the Fermi-Dirac distribution is
\begin{equation} \label{fd}
    f(\varepsilon_i) = \dfrac{1}{e^{\beta(\varepsilon_i-\mu)}+1} ,
\end{equation}
where $\mu$ is chemical potential and $\beta\equiv 1/(k_BT)$. Equation (\ref{fd}) gives the occupation probability of $\varepsilon_i$ with degeneracy $d_i$, having the average particle number
\begin{equation} \label{n}
    n_i = \dfrac{d_i}{e^{\beta(\varepsilon_i-\mu)}+1} .
\end{equation}
For systems with a conserved total particle number
\begin{equation}\label{N}
    N = \sum_i n_i = \sum_i \dfrac{d_i}{e^{\beta(\varepsilon_i-\mu)}+1} ,
\end{equation}
Equation (\ref{N}) provides a constraint from which one can solve for $\mu$ as a function of temperature $T$ and $N$. Substituting $\mu(T,N)$ to Eq. (\ref{fd}), we know how  average occupation number for each level varies with $T$ and $N$, as well as the total energy
\begin{equation}\label{E}
    E = \sum_i n_i\varepsilon_i = \sum_i \dfrac{d_i\varepsilon_i}{e^{\beta(\varepsilon_i-\mu)}+1} .
\end{equation}
%where degeneracy $d_j=2j_i+1$ for each $i$-th level of a $j$-shell is made explicit.

Now we discuss our method. For the distribution of $M$ derived in Eq. (\ref{PM}), our goal is to {\it calculate} the variance of $M$ in Eq. (\ref{sigma2}). For a nucleus with $N$ nucleons, one must know how to obtain total angular momentum $J$ given the $m$'s in Eq. (\ref{TotalM}). This nontrivial question is the same as that in the $m$-scheme shell model \cite{TalmiBook}. Our recent work \cite{Guo2025} has already provided a solution, in which we proposed a novel angular momentum coupling method using the properties of Wigner rotation theory \cite{GuidryBook}.

Our model space consists of multiple spherical $j$-shells, each characterized by single-particle energy $\varepsilon_i$, angular momentum $j_i$, with degeneracy $2j_i+1$. All $j_i$ values are assumed to be half-integers.
In Ref. \cite{Guo2025}, we distinguish the roles of single-$j$ and multi-$j$ shells in angular momentum coupling. We demonstrated that angular momentum coupling between different $j$-shells can be regarded as if they were distinguishable particles, meaning that there is no `Pauli exclusion principle between shells'. From a statistical point of view, this means that when choosing two nucleons each from different $j$-shells, their $m$ values are unrestricted. In this sense, the nucleons are independent. In statistics, the variance of independent random variables is equal to the sum of all individual variances, i.e.
\begin{equation} \label{VarM1}
    \mathrm{Var}(M) = \sum_i \mathrm{Var}(M_i). 
\end{equation}
In this way, Eq. (\ref{VarM1}) conveniently decomposes the total variance into variances of individual $j$ shells.

However, if the two nucleons are randomly selected from the same $j$ shell, their $m$ values are not allowed to be the same. Let us discuss this important case in detail.

First, consider one puts one nucleon in $j_i$-shell. Due to the $2j_i+1$ degeneracy, it is equally probable for this nucleon to be put in any $m$-state. So $m$ is uniformly distributed from $-j_i$ to $j_i$, and the variance of $m$ is
\begin{equation}\label{varm}
\mathrm{Var}(m) = \dfrac{1}{2j_i+1}\sum_{-j_i}^{j_i} m^2 = \dfrac{1}{3}j_i(j_i+1).
\end{equation}

Next, suppose that one puts $n_i$ nucleons in one same $j_i$-shell. The total $M_i$ in this shell is 
 \begin{equation}
 M_i = m_1+m_2+m_3 + \cdots m_{n_i} .
 \end{equation}
Due to the Pauli exclusion principle, no repeated $m$ values are allowed in this $j_i$-shell. As a result, the variance of $M_i$ is not equal to $n_i$ times the variance of $m$, i.e. 
\begin{equation}
\mathrm{Var}(M_i)\ne n_i\mathrm{Var}(m) .
\end{equation}
Statistically, this corresponds to {\it sampling without replacement}. Physically, the constraints imposed by fermionic antisymmetry on allowed total $J$ states are fully incorporated into the single-$j$ space.

To get $\mathrm{Var}(M_i)$, the statistical treatment is to introduce a so-called finite population correction (FPC) \cite{CochranBook} to the term  $n_i\mathrm{Var}(m)$ (see Supplemental Materials). Note that the expectation of $M_i$ still vanishes, i.e. 
\begin{equation}
\mathbb{E} [M_i] = n_i ~\mathbb{E} [m] = 0 .
\end{equation}
 The result with inclusion of the FPC factor is
\begin{equation}
\begin{aligned} \label{VarM_i}
    \mathrm{Var}(M_i) &= n_i\mathrm{Var}(m) \times \dfrac{d_i-n_i}{d_i-1} \\
    &= n_i\mathrm{Var}(m) \times \dfrac{2j_i+1-n_i}{2j_i}\\
    &= \dfrac{1}{6}n_i(j_i+1)(2j_i+1-n_i) .
\end{aligned}
\end{equation}
It is clear that $\mathrm{Var}(M_i)$ in (\ref{VarM_i}) respects the particle-hole symmetry discussed in Ref. \cite{Guo2025}: It has the same value for $n_i=n_0$ and $n_i=2j_i+1-n_0$, for any $n_0$ ($0\le n_0 \le 2j_i+1$).
The summed variance of Eq.  (\ref{VarM1}) becomes
\begin{equation}\label{VarM2}
\begin{aligned}
    \mathrm{Var}(M) &= \dfrac{1}{6}  \sum_i n_i(2j_i+1-n_i)(j_i+1) \\
    &=  \dfrac{1}{6}\sum_i \dfrac{e^{\beta(\varepsilon_i - \mu)}}{\left[e^{\beta(\varepsilon_i-\mu)}+1\right]^2} (j_i+1)(2j_i+1)^2 .
\end{aligned}
\end{equation}
In the last step, we have substituted the average particle number $n_i$ of Eq. (\ref{n}) into Eq. (\ref{VarM_i}).

Considering that statistically, protons and neutrons are two independent Fermi gases in a nucleus, the variance of the total $M$ distribution, or the squared spin-cutoff parameter $\sigma_{tot}^2$, is the sum of those from protons and neutrons, i.e.
\[
\sigma_{tot}^2 = \sigma_n^2 + \sigma_p^2 ,
\]
with $\sigma_n^2=\mathrm{Var}(M_n)$ and $\sigma_p^2=\mathrm{Var}(M_p)$ given in Eq. (\ref{VarM2}). This is the analytic form of the squared spin-cufoff parameter as function of temperature $T$ (or total energy $E$, Eq. (\ref{E})), constrained by total particle number $N$ (Eq. (\ref{N})). A set of degenerated single-particle energies $\varepsilon_i$ with angular momentum $j_i$ provide the only input for the calculation (see Supplemental Materials).

\vspace{0.2cm}
\textit{Discussion} -- Our method differs from Bethe \cite{Bethe1936} in the treatment of sample variables (i.e. the nucleons). Bethe treated the nucleons as independent, structureless particles moving in an average potential, filling the available states following the Pauli principle until reaching the Fermi level. In contrast, we strictly adhere to nuclear shell model concept, considering the nucleons moving in a realistic potential with the spin-orbit interactions \cite{RingBook} (through the use of realistic single $j$-shell energies). %which are obtained either experimentally \cite{Dyszel2025} or calculated using advanced shell models \cite{Otsuka2010}. The key to solving the problem of exponentially large configuration spaces lies in our previous work \cite{Guo2025}, which can effectively handle the angular momentum coupling of multi-$j$ shell model spaces.

\begin{figure}[H]
  \centering
  \includegraphics[width=0.45\textwidth]{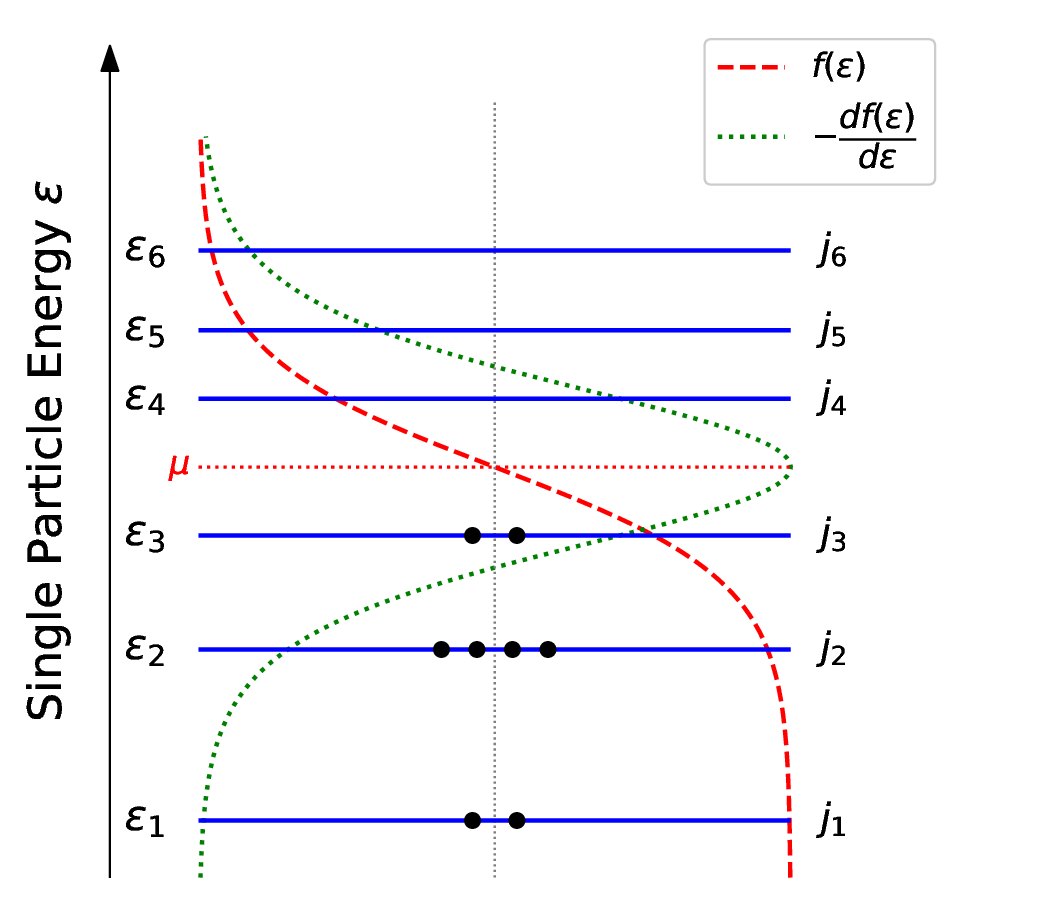}
  \caption{(Color online) Schematic diagram illustrating single-particle levels $\varepsilon_i$ with degeneracy $d_i$, chemical potential $\mu$, Fermi-Dirac distribution $f(\varepsilon)$, and its derivative $-{{df(\varepsilon)}\over{d\varepsilon}}$.}
  \label{Sn1}
\end{figure}

Why the BE formula does not include the FPC term, yet it remains valid, at least when temperature is high. This is because at the high temperature limit, the degeneracy $d$ is much greater than the occupation number $n$, so the FPC term is approximately one and can be ignored, as seen clearly in Eq. (\ref{VarM_i}). However, the FPC term typically varies with temperature-dependent occupation. To see this clearly, we rewrite Eq. (\ref{VarM2}) as
\begin{equation}
    \mathrm{Var}(M) = \sum_i w_i v_i ,
\label{wv}
\end{equation}
with $w_i=\dfrac{e^{\beta(\varepsilon_i - \mu)}}{\left[e^{\beta(\varepsilon_i-\mu)}+1\right]^2}$ and $v_i=\dfrac{1}{6}(j_i+1)(2j_i+1)^2$. The $w_i$ term, which may be interpreted as weight of the $i$ shell, contains the structure quantity $(\varepsilon_i-\mu)$ and is $T$-dependent. The $v_i$ term depends on the size of the shell and amplifies the contribution of higher $j$ shells. For a given single-particle shell, $w$ is related to the derivative of the Fermi-Dirac distribution function $f(\varepsilon)$ through
\begin{equation}
w(\varepsilon) = -k_BT\dfrac{df(\varepsilon)}{d\varepsilon} .
\end{equation}
In Fig. \ref{Sn1}, we schematically plot a set of shell-model single-particle levels $\varepsilon_j$, along with the Fermi-Dirac distribution function $f(\varepsilon)$, its derivative $-{{df(\varepsilon)}\over{d\varepsilon}}$, and chemical potential $\mu$. It can be seen that the red-dashed curve of $f(\varepsilon)$ exhibits a typical statistical distribution for $T\neq 0$, while its derivative (the green-dotted curve) presents a bell-shaped curve centered at $\mu$. At low temperatures, the $-{{df(\varepsilon)}\over{d\varepsilon}}$ curve exhibits a narrower peak, which limits the summation in Eq. (\ref{wv}) to the $j$ shells falling within the bell-shaped curve. Depending on $T$, it is likely that only one $j$-shell is located near $\mu$. As $T$ increases, the curve extends, so the summation in Eq. (\ref{wv}) may contain multiple $j$ shells. The shells with no overlap with the bell curve do not contribute to the summation, which is understandable because those shells are either completely full or completely empty, thus do not contribute to the total angular momentum.

Note that even at $T=0$, there is a non-negligible contribution to the summation in (\ref{wv}). Although $-{{df(\varepsilon)}\over{d\varepsilon}}$ is divergent at $\varepsilon=\mu$, due to the presence of the $kT$ factor, $w(\varepsilon)$ is not. As demonstrated in Supplemental Materials, $w(\varepsilon)$ reaches a nonzero limit at $T=0$, defining the minimum of $\sigma$. 

The $j$-dependent $v$ term in Eq. (\ref{wv}) further amplifies the FPC effect by up to two orders of magnitude, making the contribution from high-$j$ shells particularly significant. Overall, this is a remarkable discovery: among all nucleons in the model space, only the subset of nucleons occupying the high-$j$ shells adjacent to $\mu$ contributes to the formation of total angular momentum through correlations governed by the coupling rule. Thus, the total variance (and spin cutoff) fully reflects how the nucleons, starting from $T=0$, become {\it  correlated} due to rotational symmetry and the Pauli exclusion.

That the subset of nucleons occupying the high-$j$ shells adjacent to $\mu$ dominates the formation of total angular momentum 
contradicts the common knowledge on spin distribution. To classify the total NLD according to $J$ values, Landau hypothesized \cite{Zelevinsky2019} that a complex nucleus is analogous to a phenomenological rigid rotor with an effective moment of inertia ${\mathcal{J}}_{\rm eff}$, possessing $2J + 1$ $K$ quantum numbers to define the orientation of $J$. Total angular momentum is then provided by collective rotation of all nucleons constituting the rotor. Following this line of thought, then finding the relationship between $\sigma$ and ${\mathcal{J}}_{\rm eff}$ has been a major research direction \cite{Grimes2016}; but as we know, the research has achieved limited success.

There are familiar cases in which a small number of fermions near the Fermi level determine the overall properties of a quantum system. The Stephens-Simon effect in rotating nuclei \cite{Stephens1972,Lee1977} shows that the Coriolis field breaks Cooper pairs of high-$j$ nucleons near $\mu$ and induces their spins to align with the rotational axis, resulting in a sudden increase in collective moment of inertia and a characteristic backbending phenomenon \cite{RingBook}. %A recent proposal indicates that with  sufficiently fast nuclear rotation, all paired nucleons at the ground state can coherently align their spins along the rotation, thus breaking completely the time-reversal symmetry and forming observable topological effect \cite{Guidry2026} 

In Supplemental Materials, we show calculated $\sigma$ for $^{48, 51, 52}$Cr by applying our method. Single particle energies are the sole input to the calculation, which are the single-particle
states (with addition of two hole-states) in shell-model calculations for the same mass region \cite{Kaneko2008}. It is worth noting that our $\sigma$ includes {\it all} high-spin states allowed in the model space that make up the high-spin tails in the spin distribution, thus truly reflecting the meaning of ``spin cutoff".

\vspace{0.2cm}
\textit{ Concluding remarks } -- The present work elucidates the physical origin of the BE formula that represents the Gaussian spin distribution for nuclear level densities \cite{Bethe1936,Ericson1960}. 
Starting with Bethe, it is generally believed that the total spin is the result of a random coupling of all nucleons in angular-momentum space, an assumption justified by the complexity of the system. In contrast, in our model, angular momentum coupling {\it organizes} numerous product states, establishing correlations between nucleon orientations in Hilbert space and confining the space to irreducible SU(2) sectors. The emerging FPC term allows for sampling without replacement in finite systems with fixed degeneracy, thereby reducing the overestimation of variance in standard nuclear statistical models. Furthermore, the angular momentum of the entire system is primarily generated by states near the Fermi surface, while contributions from internal states can be neglected. Our method naturally defines the spin-cutoff $\sigma(E,J)$, which serves as a simple observable through which this geometric organization becomes accessible.

The present discussion does not involve interactions that lead to dynamic configuration mixing. We are aware that dynamic correlations can modify the effective moment of inertia and affect details of level densities, especially in the low-energy region. However, a study of these detailed modifications is not our current focus. Our study (Ref. \cite{GuoThesis}, to be published elsewhere) shows that adding interactions (e.g. the pairing force \cite{RingBook}) to this model can alter the results, particularly in the low-energy region, but does not change the current discussion and conclusions. 
As excitation increases, the level density grows exponentially ($d\gg n$ in Eq. (\ref{VarM_i})). While individual structures disappear, {\it correlations imposed by symmetry and Hilbert-space structure (largely independent of the details of interactions)} remain unaffected, and persist in highly excited states. A recent study suggests that under extremely high excitation and high spin, heavy deformed nuclei with geometrically organized states may even undergo a transition to a topological phase \cite{Guidry2026}. 

We are happy to share our code upon request; it is a Jupyter Notebook written in Python. The code can run on any personal computer with a Python compiler installed. The calculations typically take a few seconds.

\vspace{0.2cm}
\begin{acknowledgments}
This work is supported by the National Natural Science Foundation of China (Grant No.12235003).
\end{acknowledgments}

%\bibliography{ paper }

\begin{thebibliography}{}

\bibitem{Weidenmueller2009} H. A. Weidenm\"uller and G. E. Mitchell, Rev. Mod. Phys. 81, 539 (2009).

\bibitem{HF1952} W. Hauser and H. Feshbach, Phys. Rev. {\bf 87}, 366 (1952).

\bibitem{Bohr1936} N. Bohr, Nature {\bf 137}, 344 (1936).

\bibitem{Bethe1936} H. A. Bethe, Phys. Rev. {\bf 50}, 332 (1936).

\bibitem{Griffiths1995} D. J. Griffiths, {\it Introduction to Quantum Mechanics} (Prentice Hall, New Jersey, 1995).

\bibitem{Luttinger1952} J. H. D. Jensen and J. M. Luttinger, %Angular momentum distributions in the thomas-fermi model, 
Phys. Rev. {\bf 86}, 907 (1952).

\bibitem{Bloch1954} C. Bloch, %Theory of nuclear level density, 
Phys. Rev. {\bf 93}, 1094 (1954).

\bibitem{Ericson1960} T. Ericson, Advances in Physics, {\bf 9}, 425 (1960).

\bibitem{Cameron1965} A. Gilbert and A. G. W. Cameron, Can. J. Phys. {\bf 43}, 1446 (1965).

\bibitem{Cameron1969} P. J. Brancazio and A. G. W. Cameron, Can. J. Phys. {\bf 47}, 1029 (1969).

\bibitem{Dilg1973} W. Dilg, W. Schantl, H. Vonach, and M. Uhl, Nucl. Phys. A {\bf 217}, 269 (1973).

\bibitem{Su1991} Z. Su, Z. Huang, P. He, C. Zhou, Chin. J. Nucl. Phys. {\bf 13}(2), 147 (1991).

\bibitem{Iljinov1992} A. S. Iljinov, M. V. Mebel, N. Bianchi, E. De Sanctis, C. Guaraldo, V. Lucherini, V. Muccifora, E. Polli, A. R. Reolon, and P. Rossi, Nucl. Phys. A {\bf 543}, 517 (1992).

\bibitem{Ignatyuk1993} A. V. Ignatyuk, J. L. Weil, S. Raman, and S. Kahane, Phys. Rev. C {\bf 47}, 1504 (1993).

\bibitem{Rauscher1997} T. Rauscher, F.-K. Thielemann, and K.-L. Kratz, Phys. Rev. C {\bf 56}, 1613 (1997).

\bibitem{Egidy2009} T. von Egidy and D. Bucurescu, Phys. Rev. C {\bf 80}, 054310 (2009).

\bibitem{Grimes2016} S. M. Grimes, A. V. Voinov, and T. N. Massey, Phys. Rev. C {\bf 94}, 014308 (2016).

\bibitem{Voinov2023} A. V. Voinov {\it et al.} Phys. Rev. C {\bf 108}, 034302 (2023).

\bibitem{Grimes2019} S. M. Grimes, T. N. Massey, and A. V. Voinov, Phys. Rev. C {\bf 99}, 064331 (2019). 

\bibitem{Johnson2001} C. W. Johnson, J.-U. Nabi, and W. E. Ormand, arXiv:nucl-th/0111068 (2001).

\bibitem{Alhassid2005} Y. Alhassid, G. F. Bertsch, L. Fang, and S. Liu, Phys. Rev. C {\bf 72}, 064326 (2005).

\bibitem{Kaneko2007} K. Kaneko and A. Schiller, Phys. Rev. C {\bf 75}, 044304 (2007).

\bibitem{Zelevinsky2016} R. Sen'kov and V. Zelevinsky, Phys. Rev. C {\bf 93}, 064304 (2016).

\bibitem{Dang2017} N. Q. Hung, N. D. Dang, and L. T. Q. Huong, Phys. Rev. Lett. {\bf 118}, 022502 (2017).

\bibitem{Ormand2020} W. E. Ormand and B. A. Brown, Phys Rev C {\bf 102}, 014315 (2020).

\bibitem{Sun2016} Y. Sun, Phys. Scr. {\bf 91}, 043005 (2016).

\bibitem{Alhassid2007} Y. Alhassid, S. Liu, and H. Nakada, Phys. Rev. Lett. {\bf 99}, 162504 (2007).

\bibitem{Wang2023} J.-Q. Wang, S. Dutta, L.-J. Wang, and Y. Sun, Phys. Rev. C {\bf 108}, 034309 (2023).

\bibitem{Wang2025} J.-Q. Wang, S. Dutta, C.-J. Lv, L.-J. Wang, and Y. Sun, Phys. Rev. C {\bf 111}, 034324 (2025).


\bibitem{Liddick2016} S. N. Liddick {\it et al.}, Phys. Rev. Lett. {\bf 116}, 242502 (2016).

\bibitem{Stetcu2014} I. Stetcu, P. Talou, T. Kawano, and M. Jandel, Phys. Rev. C {\bf 90}, 024617 (2014).

\bibitem{Stetcu2013} I. Stetcu, P. Talou, T. Kawano, and M. Jandel, Phys. Rev. C {\bf 88}, 044603 (2013).

\bibitem{Bulgac2021} A. Bulgac, I. Abdurrahman, S. Jin, K. Godbey, N. Schunck, and I. Stetcu, Phys. Rev. Lett. {\bf 126}, 142502 (2021).

\bibitem{Schunck2021} P. Marev\'ic, N. Schunck, J. Randrup, and R. Vogt, Phys. Rev. C {\bf 104}, L021601 (2021).

\bibitem{Zeiser2019} F. Zeiser, G. M. Tveten, G. Potel, A. C. Larsen, M. Guttormsen, T. A. Laplace, S. Siem, D. L. Bleuel, B. L. Goldblum, L. A. Bernstein, F. L. Bello Garrote, L. Crespo Campo, T. K. Eriksen, A. G\"orgen, K. Hadynska-Klek, V. W. Ingeberg, J. E. Midtbø, E. Sahin, T. Tornyi, A. Voinov, M. Wiedeking, and J. Wilson, Phys. Rev. C {\bf 100}, 024305 (2019).

\bibitem{Liu2024} L.-L. Liu, Y.-Y. Liu, X.-L. Huang, and 
J.-M. Wang, Phys Rev C {\bf 109}, 014603 (2024).


\bibitem{Guo2025} J.-C. Guo and Y. Sun, %New angular momentum coupling method based on Wigner rotation theory, 
Phys. Rev. C {\bf 112}, 064307 (2025).

\bibitem{TalmiBook} A. de-Shalit and I. Talmi, {\it Nuclear Shell Theory} (Academic Press, New York and London, 1963).

\bibitem{GuidryBook} M. Guidry and Y. Sun, {\it Symmetry, Broken Symmetry, and Topology in Modern Physics: A First Course} (Cambridge University Press, Cambridge, 2022).

\bibitem{CochranBook} W. G. Cochran, {\it Sampling Techniques} (Third Ed.) (Wiley, New York, 1977).

\bibitem{RingBook} P. Ring and P. Schuck, {\it The nuclear many-body problem} (Springer Verlag, New York, 1980).

\bibitem{Zelevinsky2019} V. Zelevinsky and M. Horoi, Prog. Part. Nucl. Phys. {\bf 105}, 180 (2019).

\bibitem{Stephens1972} F. S. Stephens and R. S. Simon Nucl. Phys. A {\bf 183}, 257 (1972).

\bibitem{Lee1977} I. Y. Lee, M. M. Aleonard, M. A. Deleplanque, Y. EI-Masri, J. O. Newton {\it et al.}, Phys. Rev. Lett. {\bf 38}, 1454 (1977).

\bibitem{Kaneko2008} K. Kaneko, Y. Sun, M. Hasegawa, and T. Mizusaki, Phys. Rev. C {\bf 78}, 064312 (2008).
 
\bibitem{GuoThesis} J.-C. Guo, {\it A novel method for studying nuclear level density
based on fundamental principles of quantum mechanics}, Doctoral dissertation, Shanghai Jiaotong University, 2025.

\bibitem{Guidry2026} M. Guidry and Y. Sun, Phys. Rev. Lett. {\bf 136}, 062502 (2026).

\end{thebibliography}

{}
\end{document}